\documentclass[leqno,12pt]{article}
\usepackage[hang]{subfigure}
\usepackage{color}
\usepackage{latexsym}
\renewcommand{\epsilon}{\varepsilon}

\usepackage{amsmath}
\usepackage{amssymb}
\usepackage{ntheorem}
\usepackage[ansinew]{inputenc}
\usepackage{graphicx}
\usepackage{epsfig}
\usepackage{dsfont}
\usepackage{bm}
\usepackage{bbm}

\usepackage[authoryear]{natbib}
\newtheorem{theorem}{Theorem}[section]
\newtheorem{example}{Example}[section]

\def\3{\ss}

\newcommand{\CCp}{\mathcal{P}}
\newcommand{\CCs}{\mathcal{S}}

\newcommand{\bea}{\begin{eqnarray*}}
\newcommand{\eea}{\end{eqnarray*}}
\newcommand{\be}{\begin{eqnarray}}
\newcommand{\ee}{\end{eqnarray}}

\newcommand{\ba}{\begin{array}}
\newcommand{\ea}{\end{array}}
\def\3{\ss}

\begin{document}

\title{Optimal designs for nonlinear regression models with respect to non-informative priors}

\author{
{\small Ina Burghaus} \\
{\small Ruhr-Universit\"at Bochum} \\
{\small Abteilung f\"ur Medizinische Informatik,}\\
{\small Biometrie und Epidemiologie  } \\
{\small 44780 Bochum, Germany} \\
{\small e-mail: ina.burghaus@rub.de}\\
\and
{\small Holger Dette} \\
{\small Ruhr-Universit\"at Bochum} \\
{\small Fakult\"at f\"ur Mathematik} \\
{\small 44780 Bochum, Germany} \\
{\small email: holger.dette@rub.de}\\
}

 \maketitle

\begin{abstract}
In nonlinear regression models the Fisher information depends on the parameters of the model. Consequently, optimal designs maximizing some functional of the information matrix cannot be implemented directly but require some preliminary knowledge about the unknown parameters. Bayesian optimality criteria provide an attractive solution to this problem. These criteria depend sensitively on a reasonable specification of a prior distribution for the model parameters which might not be available in all applications. In this paper we investigate Bayesian optimality criteria with  non-informative prior distributions. In particular, we study the Jeffreys and the Berger-Bernardo  prior for which the corresponding optimality criteria are not necessarily concave.
Several examples are investigated where optimal designs with respect to the new criteria are calculated and compared to Bayesian optimal designs based on a uniform and a functional uniform prior.
\end{abstract}


\section{Introduction}\label{sec1}
\def\theequation{1.\arabic{equation}}
\setcounter{equation}{0}

Nonlinear regression models provide an important tool to describe the relation between a response and a predictor and have many applications in engineering, physics, biology, economics and medicine, among others [see \cite{ratkowsky1983}]. It is well known that a good design can improve the accuracy of the statistical analysis substantially and numerous authors have worked on the problem of constructing optimal designs for nonlinear regression models. An intrinsic difficulty of these optimization problems consists in the fact that the Fisher information, say $I(x,  \bm    {\theta})$, at an experimental condition $x$ depends on the unknown parameter $\bm{\theta} \in \bm{\Theta}$ of the model. A common approach in the literature is to assume some prior knowledge of the unknown parameter, which can be used for the construction of   optimal designs. \cite{chernoff1953} proposed the concept of local  optimality   where a fixed value of the unknown parameter is specified, and a design is determined by maximizing a functional of the information matrix for this specified parameter.

 Since this pioneering work numerous authors have constructed locally optimal designs for various regression models [see \cite{hestusun1996}, \cite{khumuksingho2006}, \cite{fangheda2008}, \cite{yangstuf2009}, \cite{yang2010} and \cite{detmel2011}, among many others]. On the other hand,
the concept of local  optimality   has been criticized by several authors, because it depends sensitively on a precise specification of the unknown parameters and can lead to inefficient designs if these parameters are misspecified [see for example \cite{detmelshp2013}, Example 2.1]. As a robust alternative \cite{pronwalt1985} and \cite{chalar1989}     proposed Bayesian optimal designs which maximize an expectation of the information criterion with respect to a prior distribution for the unknown parameters [see also \cite{chaver1995} for a review].
 Bayesian optimal designs for various prior distributions have been discussed by numerous authors [see \cite{haines1995}, \cite{detneu1997}, \cite{hancha2003} or \cite{detbra2007} among others]. However, there exist many applications where the specification of a prior distribution is difficult and
  several authors advocate  the use of a uniform prior as a pragmatic approach if no preliminary knowledge about the unknown parameter is available. In a recent paper it was pointed out by \cite{bornkamp2012} that for several models the use of a uniform prior as a non-informative prior does not yield   reasonable designs. This author proposed the concept of a functional uniform prior in order to construct Bayesian optimality criteria with respect to non-informative prior distributions.

In this paper we consider two  alternative criteria for the construction of Bayesian optimal designs with respect to non-informative prior distributions. Roughly speaking, the criteria maximize the predicted Kullback-Leibler distance between the prior and the posterior distribution for the unknown parameter of the model with respect to the choice of the experimental design, where -- in contrast to the classical approach to Bayesian optimality -- the prior distribution depends also on the design of experiment. The criteria are introduced in Section \ref{sec2}, which also gives an introduction into the field of optimal experimental design. Here it is  demonstrated that Bayesian  optimal design problems  corresponding to non-informative priors are in general not convex. Necessary conditions for the optimality of a given design are also
derived. In Section \ref{sec4} we use the theory of canonical moments which is introduced in Section \ref{sec3} [see also \cite{dettstud1997}] in order to determine saturated Bayesian optimal designs with respect to non-informative priors for polynomial regression models with a heteroscedastic error structure. Finally, in Section \ref{sec5} we consider two frequently used nonlinear regression models and compare the optimal designs with respect to the new criteria proposed in this paper with optimal designs with respect to ``classical'' Bayesian optimality criteria based on a uniform and a functional uniform distribution.

\section{Optimal design and non-informative priors} \label{sec2}
\def\theequation{2.\arabic{equation}}
\setcounter{equation}{0}

An approximate  design is defined as a probability measure $\xi$ on the design space ${\cal X}$
 with finite support [see \cite{kiefer1974}]. If the design $\xi$ has masses $\xi_i$ at the points $x_i $  $(i = 1, \dots, m)$  and $N$
observations can be made by the experimenter, this means that the quantities $\xi_i N$ are rounded to integers, say $n_i$, satisfying
$\sum^m_{i=1} N_i =N$, and the experimenter takes $N_i$ observations at each location $x_i$  $(i=1, \dots, m)$.
The corresponding design with masses $N_i/N$ at the points $x_i$  $(i=1, \ldots , m)$ will be denoted as exact design $\xi_N$.
Assume that $\xi_N $ is an  exact design with masses $N_i/N$ at points $x_i \ (i=1,\dots,m)$ and that $N_i$ independent
observations $Y_{i1},\dots,Y_{iN_i}$ are taken at each $x_i$ with density
\be \label{d1}
p(y_{ij}|\bm{\theta}, x_i) \ ; \quad j=1,\dots,N_i,\ i=1,\dots,m;
\ee
such that
\be \label{d2}
\lim_{N \to \infty} \frac {N_i}{N} = \xi_i > 0, \qquad i=1,\dots,m,
\ee
where $\bm{\theta} \in \Theta $ is a $k$-dimensional parameter. If $\xi_N$ denotes the design with masses $N_i/N$ at $x_i \ (i=1,\dots,m)$ we define by
$$
p(\bm{y}| \bm{\theta}, \xi_N) = \prod^m_{i=1} \prod^{N_i}_{j=1} p(y_{ij}|\bm{\theta}, x_i)
$$
 the joint density of the $N$-dimensional vector $\bm{Y}=(Y_{11},\dots,Y_{m N_m})^T$.
In the following we assume that the prior distribution for the parameter $\bm{\theta} $ may depend on the design
(such as the Jeffreys prior) and consider the problem of maximizing the expected Kullback-Leibler distance between the prior and posterior distribution with respect to the choice of the design $\xi_N$, that is
\be \label{d3}
    U(\xi_N) = \int \log (\frac {p(\bm{\theta}  | \bm{y},\xi_N)}{p(\bm{\theta}  | \xi_N) }) \cdot p(\bm{y},\bm{\theta}  | \xi_N) \ d\bm{\theta}  dy.
\ee
Here $
p(\bm{\theta}  |\xi_N)$      denotes the  density of the  prior distribution of $\bm{\theta} $,  $ p(\bm{\theta}  |  \bm{y}, \xi_N)  $
the density of  the posterior distribution of $\bm{\theta} $ given $\bm{y}$  and $p(\bm{y}, \bm{\theta}  | \xi_N)$ is the density of the joint distribution of  $(\bm{Y},\bm{\theta} )$.
Note that all distributions may depend on the design $\xi_N$.

Under regularity assumptions it can be shown by similar arguments as in  \citet{chaver1995}
that the expected Kullback-Leibler distance can be approximated by
\be \label{d4}
    U(\xi_N) & \approx &
  - \frac {k}{2}  \log(2\pi)- \frac {k}{2} + \frac {1}{2}  \int \log \Bigl(|N M ( \bm{ \theta} , \xi_N  | \Bigr) p(\bm{\theta} | \xi_N) \ d\bm{\theta}  \\
 \nonumber &&  -  \int \log p( \bm{ \theta}  | \xi_N) \ p(\bm{\theta}|\xi_N) \ d\bm{\theta},
\ee
where
\be \label{d5}
M(\xi_N,\bm{\theta} ) = \int \Bigl( \frac {\partial}{\partial \bm{\theta} } \log p (\bm{y}| \bm{\theta}, x) \Bigr)
\Bigl( \frac {\partial}{\partial \bm{\theta} } \log p (\bm{y}| \bm{\theta}, x) \Bigr)^T p(\bm{y}|\bm{\theta}, x)dy \xi_N (dx)
\ee
denotes the Fisher information matrix.
If the prior distribution of $\bm{\theta}$ does not depend on the design, then the criterion for Bayesian D-optimality arises, i.e.

\begin{equation} \label{bayd}
\Phi_D(\xi)=\int \log(|M(\xi,\bm{\theta})|)p(\bm{\theta})d\bm{\theta}.
\end{equation}
We call the designs maximizing the criterion \eqref{bayd} {\it Bayesian $D$-optimal designs with respect to the prior $p$}. A noninformative prior often used in applications is the uniform prior, i.e.
\begin{equation} \label{puni}
  p_{\rm uni} (\bm{\theta})\propto 1.
\end{equation}
\cite{bornkamp2012} pointed out some deficits of this prior and proposed Bayesian $D$-optimal designs with respect to functional uniform priors
\begin{equation} \label{fununi}
  p_{\rm funct } (\bm{\theta}) =  \frac{ \int_{\cal X}  | M (\delta_x,  \bm{\theta}) | ^{1/2} dx}{\int\int_{\cal X}  | M (\delta_x,  \bm{\theta}) |^{1/2} dx d \bm{\theta} },
\end{equation}
where here and throughout this paper  $\delta_x$ denotes the Dirac measure at the point $x\in {\cal X}$.
 As stated in \cite{chaver1995} a necessary and sufficient condition for Bayesian D-optimality is given by the following theorem.
 \begin{theorem} \label{thmfunct}
A  design $\xi^*$ is Bayesian D-optimal if and only if  the inequality
\be
 \int  {\rm tr} \bigl\{M^{-1} (\xi^*,\bm{\theta})  M(\xi_x,\bm{\theta})\bigr \} p(\bm{\theta}) d\bm{\theta} \leq k
\ee
 holds for all $x \in \mathcal{X}$. Moreover, there is equality for all support points of the design $\xi^*$.
 \end{theorem}

In the context of Bayesian analysis priors depending on the design are frequently  used. A typical example is the Jeffreys prior [see \cite{jeffrey1946}]
\begin{equation} \label{eq:jeff3}
p^J(\bm{\theta}  | \xi) = \frac {| M ( \bm{\theta} , \xi) |^{1/2}}{\int | M ( \bm{t} , \xi) |^{1/2}d\bm{t}} \approx \frac {| NM ( \bm{\theta} , \xi_N )  |^{1/2}}{\int |N M ( \bm{t} , \xi_N  )|^{1/2}d\bm{t}}.
\end{equation}
Using the Jeffreys prior the expression \eqref{d4} reduces to
$$
U(\xi_N) \approx V(\xi) =    - \frac {k}{2} \log (2 \pi) - \frac {k}{2} +  \log (\int |N M (\xi, \bm{\theta}   ) |^{1/2} d\bm{\theta})   .
$$
Consequently,  we call an approximate  design $\xi$ {\it Bayesian optimal with respect to the Jeffreys prior} if $\xi$ maximizes
the functional
\be \label{d8}
\Phi_J (\xi) = \int | M(\xi,\bm{\theta})|^{1/2} d\bm{\theta},
\ee
where we assume throughout this paper that the integral in \eqref{d8} is finite for all approximate designs
(sufficient for this property are compactness assumptions regarding the parameter space and continuity of the information matrix with respect to the parameter).
This criterion for the choice of an experimental design has been sporadically discussed in the literature before [see \cite{polson1992}  or \cite{firth1997a,firth1997b}].

 An intrinsic difficulty in these optimization problems consists in the fact that the criterion $\Phi_J$   is in general not convex. Consequently, standard optimal design theory based on convex optimization is not directly applicable. Nevertheless, the following results provide a necessary condition for optimality with respect to this criterion. A proof   can be found in \cite{firth1997b}.
 \begin{theorem} \label{thm1}
 If a design $\xi^*$ is Bayesian optimal with respect to the Jeffreys prior,  then the inequality
 $$
 \int  {\rm tr} \bigl\{M^{-1} (\xi^*,\bm{\theta})  M(\xi_x,\bm{\theta})\bigr \} | M(\xi^*,\bm{\theta})|^{1/2} d\bm{\theta} \leq k \int | M(\xi^*,\bm{\theta})|^{1/2} d\bm{\theta}
 $$
 holds for all $x  \in \mathcal{X}$. Moreover, there is equality for all support points of the optimal design $\xi^*$.
 \end{theorem}

The next Bayesian optimality criterion with respect to a  non-informative prior distribution is motivated by the fact that not all  components of the vector $\bm{\theta}$ are of equal importance. To be precise, we
use similar arguments as in  \cite{berger1992} and  decompose
the parameter  $\bm{\theta}$ into
 $\bm{\theta}=(\bm{\theta}^T_1,\bm{\theta}^T_2)^T$ where $\bm{\theta_1}$ and $\bm{\theta_2}$ are $k_1$ and $k_2$-dimensional parameters, respectively, and $k=k_1+k_2$. The information matrix $M(\xi,\bm{\theta})$ is decomposed in a similar way, that is
\be \label{part}
M(\xi,\bm{\theta})= \left(
\begin{array}{cc}
M_{11}(\xi,\bm{\theta}) & M_{12}(\xi,\bm{\theta}) \\
M_{21}(\xi,\bm{\theta}) & M_{22}(\xi,\bm{\theta})
\end{array}
\right)~,
\ee
where $M_{ij}(\xi,\bm{\theta}) \in \mathbb{R}^{k_i \times k_j} \ (i,j=1,2)$. In the following we assume that $\bm{\theta_2}$ is a nuisance parameter and that the parameter $\bm{\theta_1}$ is of primary interest to the
 experimenter.

This approach results in a criterion where the marginal expected Kullback-Leibler distance between the prior and posterior distribution of the parameter of primary interest $\bm{\theta_1}$ is maximized with respect to the choice of the experimental design $\xi_N$, that is
  \begin{eqnarray*}
 U_1(\xi_N) &=&\int\int \log(\frac{p(\bm{\theta_1}|\bm{y},\xi_N)}{p(\bm{\theta_1}|\xi_N)})p(\bm{\theta_1},\bm{y}|\xi_N)d\bm{\theta_1}d\bm{y}\\
 &=&\int\int \log(\frac{p(\bm{\theta}|\bm{y},\xi_N)}{p(\bm{\theta}|\xi_N)})p(\bm{\theta},\bm{y}|\xi_N)d\bm{\theta}d\bm{y}\\
 &&-\int\int \log(\frac{p(\bm{\theta_2}|\bm{\theta_1},\bm{y},\xi_N)}{p(\bm{\theta_2}|\bm{\theta_1},\xi_N)})p(\bm{\theta},\bm{y}|\xi_N)d\bm{\theta}d\bm{y}.
 \end{eqnarray*}
Under regularity assumptions it can be shown that the marginal expected Kullback-Leibler distance can be approximated by
   \begin{eqnarray}\label{du1}
 U_1(\xi_N)  &\approx&  \frac {1}{2}  \int \log \Bigl(\exp \Bigl \{ \int p(\bm{\theta_2} | \bm{\theta_1}, \xi_N) \log( \frac{ | M (\xi_N,\bm{\theta})|^{1/2}}{|M_{22}(\xi_N,\bm{\theta})|^{1/2} })
 d \bm{\theta_2}\Bigr \}\Bigr) p(\bm{\theta_1} | \xi_n) \ d\bm{\theta_1}  \\
&&  +\frac {k_1}{2}  \log(\frac{N}{2\pi e}) -  \int \log p( \bm{ \theta_1}  | \xi_N) \ p(\bm{\theta_1}|\xi_N)  \ d\bm{\theta_1}. \nonumber
  \end{eqnarray}

This follows by similar arguments as in equation \eqref{d2} [see \cite{berger1992} or \cite{ghosh1992}].

Following \cite{berger1992} we decompose the prior for the parameter $\bm{\theta}$  (which may depend on the experimental design) as
 $$
 p(\bm{\theta} | \xi_N) = p(\bm{\theta_2} |  \bm{\theta_1},\xi_N) p(\bm{\theta_1} | \xi_N),
 $$
 where $p(\bm{\theta_2} |  \bm{\theta_1},\xi_N)$ denotes the conditional density of the distribution of $\bm{\theta_2}$ given $\bm{\theta_1}$ and $p(\bm{\theta_1}|\xi_N)$ is the
 density of the prior  distribution for  $\bm{\theta_1}$. More precisely, for the conditional density of $\bm{\theta_2}$ given $\bm{\theta_1}$  an analogue of the Jeffreys prior is used, that is
 \be \label{d9}
 p^{BB}(\bm{\theta_2} | \bm{\theta_1}, \xi_N)= \frac {| M_{22}(\xi_N, \bm{\theta_1}, \bm{\theta_2})|^{1/2}}{\int | M_{22}(\xi_N,\bm{\theta_1},\bm{t_2})|^{1/2}d \bm{t_2}},
 \ee
 while the density of the prior distribution for $\bm{\theta_1}$ is given by
 \be \label{d10}
 p^{BB}(\bm{\theta_1} | \xi_N) = \exp \Bigl \{ \int p^{BB}(\bm{\theta_2} | \bm{\theta_1}, \xi_N) \log ( \frac{ | M (\xi_N,\bm{\theta})|^{1/2} }{  |M_{22}(\xi_N,\bm{\theta})|^{1/2}})  d \bm{\theta_2}\Bigr \} \cdot \alpha
 \ee
 where
  \be \label{d10a}
\alpha = \Bigl(\int\exp \Bigl \{ \int p^{BB}(\bm{\theta_2} | \bm{\theta_1}, \xi_N) \log (\frac{ | M (\xi_N,\bm{\theta})|^{1/2} }{ |M_{22}(\xi_N,\bm{\theta})|^{1/2} } )d \bm{\theta_2}\Bigr \} d \bm{\theta_1}\Bigr)^{-1}
 \ee
 is a normalizing constant. Since this pioneering work on the construction of reference priors, several authors have worked on this subject and we refer to the work of \cite{clarke1993} and \cite{kass1996} for a general discussion on this subject.
Combining this prior and equation \eqref{du1} yields the following optimality criterion
 \be \label{BB}
 \Phi_{BB}(\xi)  =  \int \exp \Bigl \{ \int \frac {|M_{22}(\xi,\bm{\theta_1},\bm{\theta_2})|^{1/2}}{\int |M_{22}(\xi,\bm{\theta_1},\bm{t_2})|^{1/2}d\bm{t_2}}
    \log (\frac{|M(\xi,\bm{\theta_1},\bm{\theta_2})|^{1/2} }{ | M_{22}(\xi,\bm{\theta_1},\bm{\theta_2})|^{1/2}}) d\bm{\theta_2} \Bigr \} d \bm{\theta_1},
 \ee
 where we again assume that the integral exists for all designs $\xi$.
 Designs maximizing the function $\Phi_{BB}$ are called {\it Bayesian optimal with respect to the Berger-Bernardo prior}.
Again this criterion is in general not convex and a necessary condition for optimality will be derived.

 \begin{theorem} \label{thmBB}
 If a design $\xi^*$ is Bayesian-optimal with respect to the Berger-Bernardo prior, then  the inequality
 \bea
 d (\xi^*, \delta_x) &=& \int \int {\rm tr} \bigl[ M^{-1}_{22} (\xi^*, \bm{\theta}) M_{22}(\delta_x, \bm{\theta})\bigr] \log (\frac{ |M(\xi^*,\bm{\theta})|^{1/2}
}{ | M_{22}(\xi^*,\bm{\theta})|^{1/2}}) p^{BB}(\bm{\theta}, \xi^*) d \bm{\theta} \\
 &-& \int \Bigl \{ \int  \log  { (\frac{|M(\xi^*,\bm{\theta_1},\bm{t_2})|^{1/2} }{  | M_{22}(\xi^*, \bm{\theta_1},\bm{t_2})|^{1/2}} )}
 p^{BB}  (\bm{t_2}| \bm{\theta_1}, \xi^*)d \bm{t_2} \\
 && \times  \int {\rm tr} (M^{-1}_{22} (\xi^*, \bm{\theta_1}, \bm{t_2}) M_{22} (\delta_x, \bm{\theta_1}, \bm{t_2})) p^{BB}(\bm{t_2} | \bm{\theta_1}, \xi^*) d \bm{t_2} \Bigr \} p^{BB} (\bm{\theta_1} | \xi^*) d \bm{\theta_1} \\
 &+& \int \int \bigl[{\rm tr} (M^{-1} (\xi^*, \bm{\theta}) M (\delta_x, \bm{\theta})) - {\rm tr} (M^{-1}_{22}(\xi^*, \bm{\theta}) M_{22} (\xi_x, \bm{\theta}))\bigr] p^{BB} (\bm{\theta} | \xi^*) d \bm{\theta} \leq k_1
 \eea
 holds for all $x \in \mathcal{X}$, where $p^{BB}(\bm{\theta_2} | \bm{\theta_1}, \xi)$ and $p^{BB}(\bm{\theta_1} | \xi)$ are defined by \eqref{d9} and \eqref{d10}, respectively, and $p^{BB}(\bm{\theta} | \xi)=p^{BB}(\bm{\theta_2} | \bm{\theta_1}, \xi)p^{BB}(\bm{\theta_1} | \xi)$. Moreover, there is equality for all support points of the design $\xi^*$.
 \end{theorem}

 {\bf Proof.} The proof follows by a standard argument calculating the directional derivative
 $$\frac {\partial}{\partial t} \Phi_{BB}(\xi_t)|_{t=0},$$
  where the design $\xi_t$ is defined by $\xi_t=\xi^* + t(\eta-\xi^*)$, $\eta$ denotes an additional approximative design and $t \in (0,1)$.  Observing the fact
 $$
 \frac {\partial}{\partial t} \log | M(\xi_t, \bm{\theta})| \Big|_{t=0} = {\rm tr} (M^{-1} (\xi^*, \bm{\theta}) (M(\eta,\bm{\theta}) - M(\xi^*, \theta)))
 $$
 we obtain (recalling the definitions \eqref{d9}, \eqref{d10} and \eqref{d10a})
 \bea
 &&\frac {\partial}{\partial t} \Phi_{BB} (\xi_t) \Big|_{t=0} \\
 &&  \quad = \frac {1}{2\alpha} \int p^{BB} (\bm{\theta_1} | \xi^*) \int \Bigl [ \log (\frac {|M(\xi^*, \bm{\theta})|^{1/2}}{|M_{22}(\xi^*, \bm{\theta})|^{1/2}}) \Bigl \{ p^{BB} (\bm{\theta_2}|\bm{\theta_1},\xi^*) {\rm tr} (M^{-1}_{22} (\xi^*, \bm{\theta}) (M_{22}(\eta,\bm{\theta})-M_{22}(\xi^*,\bm{\theta})) \\
 && \quad -p^{BB} (\bm{\theta_2}|\bm{\theta_1},\xi^*) \int p^{BB} (\bm{t_2} | \bm{\theta_1}, \xi^*) {\rm tr} (M^{-1}_{22} (\xi^*, \bm{\theta_1}, \bm{t_2}) (M_{22}(\eta, \bm{\theta_1},\bm{t_2})-M_{22}(\xi^*, \bm{\theta_1}, \bm{t_2})) d \bm{t_2} \Bigr \} \\
 && \quad + p^{BB} (\bm{\theta_2} | \bm{\theta_1},\xi^*) \Bigl \{ {\rm tr} (M^{-1} (\xi^*, \bm{\theta}) (M(\eta, \bm{\theta}) - M(\xi^*,\bm{\theta})))\\
 &&  \quad - {\rm tr} (M^{-1}_{22} (\xi^*, \bm{\theta}) (M_{22}(\eta, \bm{\theta}) - M_{22} (\xi^*, \bm{\theta})) \Bigr \} \Bigr ] d \bm{\theta_2} d \bm{\theta_1} \\
 && \quad = \frac {1}{2\alpha}(d (\xi, \eta) - k_1).
\eea
The assertion now follows  by the same arguments as given in \cite[p.19]{silvey1980}. \hfill$\Box$

\bigskip

In the following chapters we will discuss optimal designs maximizing the criteria \eqref{d8} and \eqref{BB} in several examples.

\section{Canonical moments} \label{sec3}
\def\theequation{3.\arabic{equation}}
\setcounter{equation}{0}
In    Section \ref{sec4} we discuss Bayesian optimal designs with respect to non-informative priors for heteroscedastic polynomial regression
models. An important tool to derive optimal saturated designs for polynomial models is the theory of canonical moments which
was firstly used by \cite{studden1980,studden1982}   to determine $D_s$-optimal designs for homoscedastic
polynomial regression explicitly and
will be briefly introduced in this section.  Since these seminal papers numerous authors have used this methodology
to determine optimal designs in polynomial and trigonometric regression models [see \cite{laustu1985}, \cite{spruill1990}, \cite{dette1994a,dette1995b},
  and \cite{zentsa2004} among many others]. A detailed description of the theory of canonical moments can be found
in the monograph of \cite{dettstud1997}. \\
To be precise let $a,b\in  \mathbb{R}$ denote two constants such that $a<b$ and introduce by  $\CCp([a,b])$   the set of all probability measures   on the interval $[a,b]$.
We define   for a design $\xi \in \CCp([a,b])$ its moments by
 $$
c_i =c_i(\xi) =\int^b_{a} x^i\xi(dx),\,\, i=1,2 \ldots  .
$$
Define  $\mathcal{M}_n = \{(c_1,\dots,c_n)^T \mid \xi \in \CCp([a,b])\}$ as the $n$th moment space and $ \Phi_{n}(x) =(x,\ldots
,x^{n})$ as the vector of   monomials of order $n$. Consider
for a fixed vector $\bm{c}_n =(c_1, \ldots , c_{n})^T \in {\cal M}_{n}$ the set
$$\CCs_{n}(\bm{c}_n)=
\Bigl\{\xi\in\CCp([a,b]):\;\int_{a}^{b}\Phi_{n}(x)\xi(dx)=\bm{c}_n \Bigr\}$$     of all probability measures on the interval $[a,b]$ whose
moments up to the order $n$ coincide with $\bm{c}_n =(c_1, \ldots , c_{n})^T$.
 For  $n = 2,3,  \ldots$ and for a given point $(c_1, \ldots ,
c_{n-1})^T \in {\cal M}_{n-1}$ we define $c^+_n = c^+_n(c_1, \ldots , c_{n-1})$ and $c^-_n = c^-_n(c_1, \ldots , c_{n-1})$ as the largest and
smallest value of $c_n$ such that $(c_1, \ldots , c_n)^T \in \partial {\cal M}_n$ (here $\partial {\cal M}_n $ denotes the boundary of ${\cal M}_n $), that is
\begin{eqnarray*}
c^{-}_n &=& \min
\Bigl\{ \int_{a}^b x^n\xi(dx)\mid \xi \in {\it S}_{n-1}(c_1, \ldots , c_{n-1})\Bigr\} ,\\
c^{+}_n &=& {\max} \Bigl\{ \int_{a}^b x^n\xi(dx)\mid \xi \in {\it S}_{n-1}(c_1, \ldots , c_{n-1})\Bigr\}.
\end{eqnarray*}
Note that $c^-_n \le c_n \le c^+_n$ and that both inequalities are strict if and only if $(c_1, \ldots , c_{n-1})^T \in {\rm int} ( \mathcal{M}_{n-1})$ where ${\rm int} (\mathcal{M}_{n-1})$ denotes the interior of the set $\mathcal{M}_{n-1}$
[see \cite{dettstud1997}].

 For a design $\xi$ on the interval $[a,b]$ with corresponding moment point $\bm{c}_n = (c_1(\xi), \ldots , c_n(\xi))^T$,  such that $\bm{c}_{n-1} = (c_1(\xi), \ldots , c_{n-1}(\xi))^T$ is
 in the interior of the moment space ${\cal M}_{n-1}$, the
canonical moments or canonical coordinates     are defined by $p_1=c_1(\xi)$ and
\begin{equation}\label{2.4}
 p_i = p_i(\xi)= \frac{c_i(\xi) - c^-_i}{c^+_i - c^-_i},
 \quad i = 2, \ldots , n \ .
\end{equation}
Note that the canonical moments $p_i$ vary independently in the interval $[0,1]$ (whenever they are defined). Moreover, it follows that   $p_i \in (0,1)$, $i=1,\ldots ,n-1$ and $p_n\in \{0,1\}$
if and only if  $(c_1(\xi),\dots,c_{n-1}(\xi)) \in {\rm int} (\mathcal{M}_{n-1})$ and $(c_1(\xi), \ldots , c_{n}(\xi))^T \in \partial {\cal M}_n $.
In this case the canonical moments   $p_i$ of order    $i>n$ remain
undefined. \\
The main idea of \cite{studden1980} was to describe designs in terms of their canonical moments, to find a (simple) representation of the optimality criterion by these quantities and to perform optimization on the
unit cube. For this purpose optimality criteria have to be expressed explicitly in terms of canonical moments and we
recall the following basic facts [for a proof see \cite{studden1982,studden1982a} and \cite{laustu1988}].
 \begin{theorem} \label{thm2a}
 Let  $\xi$  denote a design on the interval $[a,b] $ with moments $c_1, c_{2}, \ldots$, canonical moments $p_1, p_{2}, \ldots$ and $q_0=1,q_1=1-p_1,q_2=1-p_2, \ldots$
 \begin{itemize}
 \item[(a)] Let  $H_n(\xi) = (c_{i+j})_{i,j=1,\ldots ,n}$
 denote  the Hankel matrix of the moments of the design $\xi$. If if $(c_1, \ldots , c_{2n-1})^T \in \mbox{int} ( {\cal M}_{2n-1})$,   then
 $$
 |H  (\xi )| = (b-a)^{n(n+1)} \prod_{i=1}^n (q_{2i-2}p_{2i-1}q_{2i-1}p_{2i})^{n-i+1}
  .$$
   \item[(b)]  Let $\xi$ denote a design on the interval $[a,b]$ with $n+1$ support points $x_1, \ldots x_{n+1}$, then
\begin{eqnarray*}
   \prod_{i=1}^{n+1} (x_i-a) &=& (b-a)^{n+1}p_{2n+1}\prod_{i=1}^np_{2i-1}q_{2i}~,~~
     \prod_{i=1}^{n+1} (b-x_i) = (b-a)^{n+1}\prod_{i=1}^{2n+1}q_{i}, \\
   \sum_{i=1}^{n+1} (x_i -a) &=& (b-a) \sum_{i=1}^{2n+1} q_{i-1}p_i.
\end{eqnarray*}
  \end{itemize}
 \end{theorem}
 The following results are shown in \cite{dettstud1997} and can be used to derive a design  corresponding to
 an ``optimal'' sequence of canonical moments (i.e. a sequence maximizing a particular optimality criterion).
  \begin{theorem} \label{thm2}
Let  $\xi$  denote  a design on the interval $[a,b]$ with canonical moments $p_1,.p_2, \ldots $.
\begin{enumerate}
\item[(1)]  If $p_i\in (0,1)$, $i=1,\ldots , 2n-1$ and $p_{2n}=0$, then  $\xi$  has  $m=n$  support points in the interior of the interval $(a,b)$.
\item[(2)]  If $p_i\in (0,1)$, $i=1,\ldots , 2n$ and $p_{2n+1}=0$, then  $\xi$  has  $m=n+1$  support points, $n$ points   in the interior of the interval $(a,b)$
and the point $a$.
\item[(3)]  If $p_i\in (0,1)$, $i=1,\ldots , 2n$ and $p_{2n+1}=1$, then  $\xi$  has  $m=n+1$  support points, $n$ points   in the interior of the interval $(a,b)$
and the point $b$.
\item[(4)]  If $p_i\in (0,1)$, $i=1,\ldots , 2n-1$ and $p_{2n}=1$, then  $\xi$  has  $m=n+1$  support points, $n-1$ points  in the interior of the interval $(a,b)$
and the points $a$ and  $b$.
\end{enumerate}
Moreover, the support points $x_1,\dots,x_m$ are the roots of the polynomial $P_{m}(x) = W_{m}(x)$, where
the polynomials $W_i(x)$ are defined recursively by
\begin{equation} \label{rec}
 W_{i+1}=(x-a-(b-a)(\zeta_{2i}+\zeta_{2i+1}))W_i(x)-(b-a)^2\zeta_{2i-1}\zeta_{2i} W_{i-1}(x),
\end{equation}
with initial conditions $W_0(x)=1, W_{-1}(x)=0$ and we use the notation
$\zeta_{0}=0, \zeta_{1}=p_1,\zeta_{i}=(1-p_{i-1})p_i, i\ge 2$. The weights $\xi(x_1),\ldots, \xi(x_m)$
at the support points
$x_1,\ldots, x_m$ are obtained by the formula
\begin{equation} \label{gew}
 \xi(x_i)=\frac{P^{(1)}_{m-1}(x_i)}{\frac{\partial}{\partial x}P_m(x) |_{x=x_i}} ~;~~i=1,\ldots ,m,
\end{equation}
where  $ P^{(1)}_{i}(x)=W_{i+1}(x)$ and the polynomials $W_i(x) $ are defined recursively by
\eqref{rec}
with initial conditions
$W_{1}(x)=1, W_0(x)=0$.
 \end{theorem}

\section{Robust designs for  heteroscedastic polynomials}\label{sec4}
\def\theequation{4.\arabic{equation}}
\setcounter{equation}{0}

We are now in a position to determine Bayesian optimal saturated designs with respect to non-informative priors for the polynomial regression model.
To be precise, we assume that the density $p(\bm{y}|\bm{\theta},x) $ of the response $Y$ (at experimental condition $x$) is governed by a by
normal distribution with mean
\begin{equation} \label{pol}
\mu( x,\bm{\theta}) = \sum_{j=0}^{n} \theta_j x^j
\end{equation}
 and variance $ \sigma^2(x,\bm{\theta})$, where the variance and design space are given by
\begin{eqnarray}
  \label{var2}
 \sigma^2(x,\bm{\theta}) &=& \theta_{n+1} \exp(\theta_{n+2}x)~,~~{\cal X} = [0,b] ~~(\theta_{n+1}>0,\theta_{n+2} \ge 0) \\
 \label{var1}
 \sigma^2(x,\bm{\theta}) &=& (1-x)^{-\theta_{n+1}-1 }(1+x)^{-\theta_{n+2}-1}~,~~{\cal X} = (-1,1) ~~(\theta_{n+1},\theta_{n+2} > 0)
 \end{eqnarray}
 and $b>0$ is a constant. We also note  that there are several other variance functions, which are usually investigated in the context of polynomial regression [see \cite{karstu1966}, p.\ 328,
 \cite{chang2005} or \cite{chachawan2009}].
For these variance functions similar
results to those described in the following section can be obtained, but the details are omitted for the sake of brevity.\\
Adapting the notation of the previous section we have for the parameter of interest $\bm{\theta_1}=(\theta_0,\dots,\theta_n)^T$ and for the nuissance parameters $ \bm{\theta_2} = (\theta_{n+1}, \theta_{n+2})^T$. The Fisher information at a point $x\in {\cal X}$  is given by
 \be \label{fishpol}
I(x,\bm{\theta})= \left(
\begin{array}{cc}
I_{11}(x,\bm{\theta}) & I_{12}(x,\bm{\theta}) \\
I_{21}(x,\bm{\theta}) & I_{22}(x,\bm{\theta})
\end{array}
\right) \in  \mathbb{R}^{n+3\times n+3},
\ee
where $I_{12}(x,\bm{\theta}) =0 \in  \mathbb{R}^{n+1\times 2}$ and
 \begin{eqnarray} \label{fishpol1}
 I_{11}(x,\bm{\theta}) &=&  \sigma^{-2} (x,\bm{\theta}) (x^{i+j})_{i,j=0,\ldots ,n} \in  \mathbb{R}^{n+1 \times n+1} , \\
  \label{fishpol2}I_{22}(x,\bm{\theta}) &=&
 {\frac{1}{ 2 \sigma^{2} (x,\bm{\theta})}} \Bigl( {\frac{\partial  }{\partial  {\bm{\theta_2}}}}  \sigma^2(x,\bm{\theta}) \Bigr) \Bigl( {\frac{\partial }{ \partial  { \bm{\theta_2}}}}  \sigma^2(x,\bm{\theta}) \Bigr)^T \in  \mathbb{R}^{2\times 2}.
\end{eqnarray}

In the following we call a design optimal $m$-point design, if it maximizes a particular optimality criterion in the class of all designs supported at $m$ points.
Our first result describes the class of all Bayesian-optimal $(n+1)$-point designs
for polynomial regression and  variance function \eqref{var2}
with respect to the Jeffreys and the Berger-Bernardo  prior.

  \begin{theorem} \label{thm3} Consider the polynomial regression model \eqref{pol} with variance function \eqref{var2} and design space
   ${\cal X}=[0,b]$.
  \begin{itemize}
  \item[(1)] Assume that $(\theta_0,\dots,\theta_{n+2}) \in \Theta  \subset (\mathbb{R}^+_0)^{(n+1)}\times \mathbb{R}^+\times \mathbb{R}^+_0$, where $\Theta$ is a compact set. The canonical moments of the
  Bayesian optimal $(n+1)$-design with respect to the Jeffreys prior are given by $(p_1,\dots,p_{2n-1},1)$, where $p_1,\dots,p_{2n-1} \in (0,1)$ are obtained as  a
  solution of the system of equations
  \begin{flalign*}&&&0=\frac{n-i+1}{2}(\frac{1}{p_{2i-1}}-\frac{1}{q_{2i-1}})+(p_{2i}-q_{2i-2})(\sum_{j=1}^{2n}q_{j-1}p_j)^{-1}\ (i=2,...,n)&\\
&&&0=\frac{n-i+1}{2p_{2i}}-\frac{n-i}{2q_{2i}}+(p_{2i+1}-q_{2i-1})(\sum_{j=1}^{2n}q_{j-1}p_j)^{-1}\ (i=2,...,n-1)\\
&&&0=\frac{n+1}{2}(\frac{1}{p_{1}}-\frac{1}{q_{1}})+(p_{2}-1)(\sum_{j=1}^{2n}q_{j-1}p_j)^{-1}&\\
&&&0=\frac{n+1}{2p_{2}}-\frac{n-1}{2q_{2}}+(p_{3}-q_{1})(\sum_{j=1}^{2n}q_{j-1}p_j)^{-1}.&
\end{flalign*}

  \item[(2)]  Assume that $(\theta_0,\dots,\theta_{n+2}) \in \Theta \subset (\mathbb{R}^+_0)^{(n+1)}\times (\mathbb{R}^+)^{2}$ is a compact set, denote by
$z$ the largest root of the  $n$th Laguerre polynomial  $L_n^{(1)}(x)$ and define
$$
  \gamma= \frac{\int \theta_{n+2}d{{\theta}_{n+2}}}{\int d{{\theta}_{n+2}}}.$$
\begin{itemize}
\item[(a)]
If $b \gamma \ge z$,
 then the Bayesian optimal $(n+1)$-design with respect to the Berger-Bernardo  prior
puts equal masses at the roots of the polynomial $xL_n^{(1)}(x\gamma)$.
\item[(b)]
If $b \gamma < z$,  then the canonical moments of the
  Bayesian optimal $(n+1)$-design with respect to  Berger-Bernardo  prior are obtained as  a
  solution of the system of equations  $p_{2n}=1$
\begin{flalign*}
&&&\frac{n-i+1}{p_{2i-1}}-\frac{n-i+1}{1-p_{2i-1}}-b \gamma (1-p_{2i-2})+b \gamma p_{2i}=0 \ (i=1,...,n)&\\
&&&\frac{n-i+1}{p_{2i}}-\frac{n-i}{1-p_{2i}}-b \gamma (1-p_{2i-1})+b \gamma p_{2i+1}=0 \ (i=1,...,n-1)&
\end{flalign*}
with $q_0 = 0$.
Moreover, the optimal design has equal masses at its support points.
\end{itemize}
\end{itemize}
  \end{theorem}

{\bf Proof.} Note that the lower diagonal block of the Fisher information is given by
\begin{eqnarray*} I_{22} (x,{ \bm{\theta}}) =\frac{1}{2\theta_{n+1}^2}\left(
\begin{array}{cc}
1&\theta_{n+1}x\\
\theta_{n+1}x&\theta_{n+1}^2x^2
\end{array}\right) \ .
\end{eqnarray*}
If $\xi$ denotes a design with $n+1$ support points $x_1,\dots,x_{n+1}$, then it follows from Theorem \ref{thm2a} that
\begin{eqnarray}
|M_{11}(\xi,{\bm{\theta}})|&=&\frac{|H(\xi)|}{(\theta_{n+1})^{n+1}}\exp(-b\theta_{n+2}\sum_{i=1}^{n+1}x_i)\\\nonumber
&=&
(\frac{b^n}{\theta_{n+1}})^{n+1}\prod_{j=1}^n(q_{2j-2}p_{2j-1}q_{2j-1}p_{2j})^{n-j+1} \exp(-b\theta_{n+2}\sum_{j=1}^{2n+1}q_{j-1}p_j) .
\label{m11}
\end{eqnarray}
Moreover, the canonical moments $p_1$ and $p_2$ are related to the moments $c_1$ and $c_2$ by
$c_1=bp_1$ and $c_2=b^2(p_1+q_1p_2)$, respectively [see \cite{dettstud1997}], which yields for the lower right block of the matrix $M(\xi,{ \bm{\theta}})$ in \eqref{part}
\begin{eqnarray} \label{neu}
 | M_{22}(\xi,\bm{\bm{\theta}})|=\frac{1}{4}\theta_{n+1}^{-2}b^2p_1p_2q_1 .
\label{m22}
\end{eqnarray}
Consequently, Bayesian optimal designs with respect to the Jeffreys and the Bernardo-Berger prior depend only on the parameter $\theta_{n+2}$, and only this dependence will be reflected in the optimality criterion.

For a proof of (1) note that the criterion \eqref{d8} reduces to
\begin{eqnarray*}
\Phi_J (\xi)  
&=&\alpha_1\int (\frac{\theta_{n+1}^{-{\frac{n+3}{ 2} }} }{2} b^{{\frac{n(n+1)}{ 2}} +1})d\theta_{n+1} \int_{ \mathbb{R}^+}\Bigl[\prod_{j=1}^n(q_{2j-2}p_{2j-1}q_{2j-1}p_{2j})^{\frac{n-j+1}{ 2 }}\\
&&\times\exp\Bigl(-{b\theta_{n+2}\over 2} \sum_{j=1}^{2n+1}q_{j-1}p_j\Bigr) (p_1p_2q_1)^{\frac{1}{2}}\Bigr]
d\theta_{n+2} \\
&=& \frac{\alpha_2}{b\sum_{j=1}^{2n+1}q_{j-1}p_j}
 \prod_{j=1}^n(q_{2j-2}p_{2j-1}q_{2j-1}p_{2j})^{\frac{n-j+1}{2}}(p_1p_2q_1)^\frac{1}{2}
\end{eqnarray*}
with appropriate constants $\alpha_1$ and $\alpha_2$.
Obviously this expression is maximized if $(p_1,\dots,p_{2n-1}) \in (0,1)^{2n-1}$ and $q_{2n}p_{2n+1}=0$, which can be achieved either by $p_{2n+1}=0$ and $p_{2n}\in (0,1)$ or if $p_{2n}=1$.  Now assume that  $p_{2n} \in (0,1)$ then
$(p_1, \ldots p_{2n}) \in (0,1)^{2n}$ would be a solution of  the system of equations ${\partial \over \partial p_j} \log \Phi_J (\xi)   =0 $, $j=1,\ldots , 2n$. The derivative with respect
to the coordinate $p_{2n}$ yields the equation
$$
{\partial \over \partial p_{2n}} \log \Phi_J (\xi)  = \frac{1}{2p_{2n}}-\frac{q_{2n-1}}{\sum_{j=1}^{2n}q_{j-1}p_j} =0~,
$$
which gives
\begin{equation}\label{wider}
 q_{2n-1}p_{2n} =  \sum_{j=1}^{2n-1}q_{j-1}p_j .
 \end{equation}
  Inserting this expression in the partial derivative with respect to $p_{2n-1}$ yields
\begin{eqnarray*}
{\partial \over \partial p_{2n-1}} \log \Phi_J (\xi)   &=&
{{1\over p_{2n-1}} -{1\over q_{2n-1}} \over 2} - { q_{2n-2}-p_{2n} \over  \sum_{j=1}^{2n}q_{j-1}p_j} =
{{1\over p_{2n-1}} -{1\over q_{2n-1}} \over 2}  - { q_{2n-2}-p_{2n} \over  2p_{2n}q_{2n-1}} =0~,
\end{eqnarray*}
which is equivalent to  $p_{2n}q_{2n-1}=q_{2n-2}p_{2n-1}$.  Combining this equation with \eqref{wider} gives
$$\sum^{2n-2}_{j=1} q_{j-1}p_j = 0,$$
which is a contradiction to the assumption
$p_{i} \in (0,1) \ (i=1,\dots,2n)$. Consequently, we have $p_{2n}=1$ and calculating ${\partial \over \partial p_j} \log \Phi_J (\xi)   =0 $ for  $j=1,\ldots , 2n-1$
gives the system of equation  stated in part (1)  of Theorem \ref{thm3}.

We now turn to a proof of part (2). Recall  the representation \eqref{neu}, which  yields for the  first ratio of the determinants in   criterion \eqref{BB}
$$
 \frac {|M_{22}(\xi,\bm{\theta})|^{1/2}}{\int |M_{22}(\xi,\bm{\theta_1},\bm{t_2})|^{1/2}d\bm{t_2}}  ~=~{\theta_{n+1}^{-1} \over \int t_{n+1}^{-1} dt_{n+1}\int  dt_{n+2}} ={\theta_{n+1}^{-1} \over  \alpha_3},
$$
where the last equality defines the constant $\alpha_3$ in an obvious manner.  We
introduce the notation
$$
\alpha_1 = \int d\bm{\theta_1} ~;~~\alpha_2 =\int d\theta_{n+2}~;~~\alpha_4 =\int \theta_{n+2} d\theta_{n+2}.
$$
Observing the
fact that the Fisher information matrix is
block diagonal we obtain
$$|M(\xi,\bm{\theta} )| /|M_{22}(\xi,\bm{\theta} )| = |M_{11}(\xi,\bm{\theta} )|, $$  and  \eqref{m11}  yields for the optimality
criterion \eqref{BB}
\begin{eqnarray*}
 \Phi_{BB}(\xi) &=&\alpha_1 \exp\Bigl(\int\Bigl[\frac{n+1}{2\alpha_3}\theta_{n+1}^{-1}\log(\frac{1}{\theta_{n+1}})]d\bm{\theta_2}+\frac{1}{2}\bigl[n(n+1)\log(b)\\
&&~~+\sum_{j=1}^n(n-j+1)\log(q_{2j-2}p_{2j-1}q_{2j-1}p_{2j})-\frac{b\alpha_4}{\alpha_2}\sum_{j=1}^{2n+1}q_{j-1}p_j\Bigr]\Bigr).
\end{eqnarray*}
Consequently, the Bayesian optimal $(n+1)$-point design with respect to the Berger-Bernardo  prior is obtained by maximizing
the expression
$$
\sum_{j=1}^n(n-j+1)\log(q_{2j-2}p_{2j-1}q_{2j-1}p_{2j})-\frac{b\alpha_4}{\alpha_2}\sum_{j=1}^{2n+1}q_{j-1}p_j
$$
with respect to the canonical moments $(p_1,\ldots , p_{2n+1}) \in [0,1]^{2n+1}$ and identifying the design corresponding to these canonical moments
by Theorem \ref{thm2}.
But this problem has been solved by \cite{detwon1998}, and
the assertion follows from Theorem \ref{thm2} in this reference observing that $\gamma=\alpha_4/\alpha_2$.
  \hfill $\Box$

\begin{example}
{\rm In this example we illustrate the application of  Theorem \ref{thm3} by calculating Bayesian optimal designs with respect to non-informative
priors in the polynomial regression model \eqref{pol} with variance function \eqref{var2}. Recall that only the parameter $\theta_{n+2}$ appears in the optimality criterion
in a non-trivial way  and as a consequence  Bayesian optimal designs depend only on prior information regarding this parameter. We assume that  $\theta_{n+2} \in [0,4]$.
In Table \ref{tab1} we present Bayesian optimal $4$-point designs for the cubic regression model on the interval ${\cal X} =[0,1]$ with respect to the Jeffreys prior, the Berger-Bernardo prior and Bayesian $D$-optimal $4$-point design with respect to a
uniform distribution.  The Bayesian $D$-optimal design with respect to the uniform prior and Bayesian optimal design with respect to the
Bernardo-Berger prior are similar, where the latter puts less weights at the boundary of the design space. On the other hand the support
points of the Bayesian optimal design with respect to the
Jeffreys prior in the interior of the design space are larger. In Figure \ref{fig1} we illustrate the application of Theorem \ref{thm1} and \ref{thmBB}. We observe that all designs satisfy the necessary condition for optimality.\\
Corresponding results for  a quadratic polynomial regression model are depicted in Table \ref{tab2}, where the design space is now given by the interval $[0,3]$.  Here the right boundary point of the design space is a support point of the   Bayesian optimal design with respect to the Jeffreys prior and the Bayesian $D$-optimal design with respect to
the uniform prior. On the other hand the Bayesian  optimal design with respect to the Berger-Bernardo prior does not contain the point $3$ in its support.
We observe from Figure \ref{fig1b}  that not all designs satisfy the necessary condition of optimality. Therefore we maximized the criteria for $4$-point designs numerically,
and the corresponding  designs are shown in Table \ref{tab2b}. Only the  criterion based on the Jeffreys prior yields a  $3$-point design while the other  hand two criteria
yield $4$-point designs. Moreover, all  designs meet the corresponding necessary condition for optimality (these results are not depicted for the sake of brevity).

 \begin{table}[h!]
\begin{center}
  \begin{tabular}{|c|c|c|c|c|c|}\hline
  \multicolumn{2}{|c|}{\eqref{bayd} with \eqref{puni}}&\multicolumn{2}{|c|}{\eqref{d8}}&\multicolumn{2}{|c|}{\eqref{BB}}\\\hline
  $\xi(x_i)$&$x_i$	&$\xi(x_i)$&$x_i$&$\xi(x_i)$&$x_i$		\\\hline\hline
  0.2760&0&0.2809&0&0.25&0  \\\hline
  0.2195&0.2072&0.2170&0.2347&0.25 &0.2177\\\hline
  0.2082&0.6606&0.2114&0.7018&0.25&0.6497 \\\hline
  0.2963&1&0.2907&1&0.25&1	\\\hline
  \end{tabular}
 \caption{ \label{tab1} \it Bayesian optimal $4$-point designs with respect to non-informative priors  for the
 cubic polynomial regression model on the interval $[0,1]$  with variance structure  \eqref{var2}, where
 $\theta_{n+2} \in [0,4]$. Left column: Bayesian $D$-optimal designs with respect to the uniform prior. Middle column: Bayesian optimal designs
 with respect to the Jeffreys prior. Right  column: Bayesian optimal designs
 with respect to the Bernardo-Berger prior.}

 \end{center}
 \end{table}

   \begin{figure}[hbt!]
     \includegraphics[width=5cm]{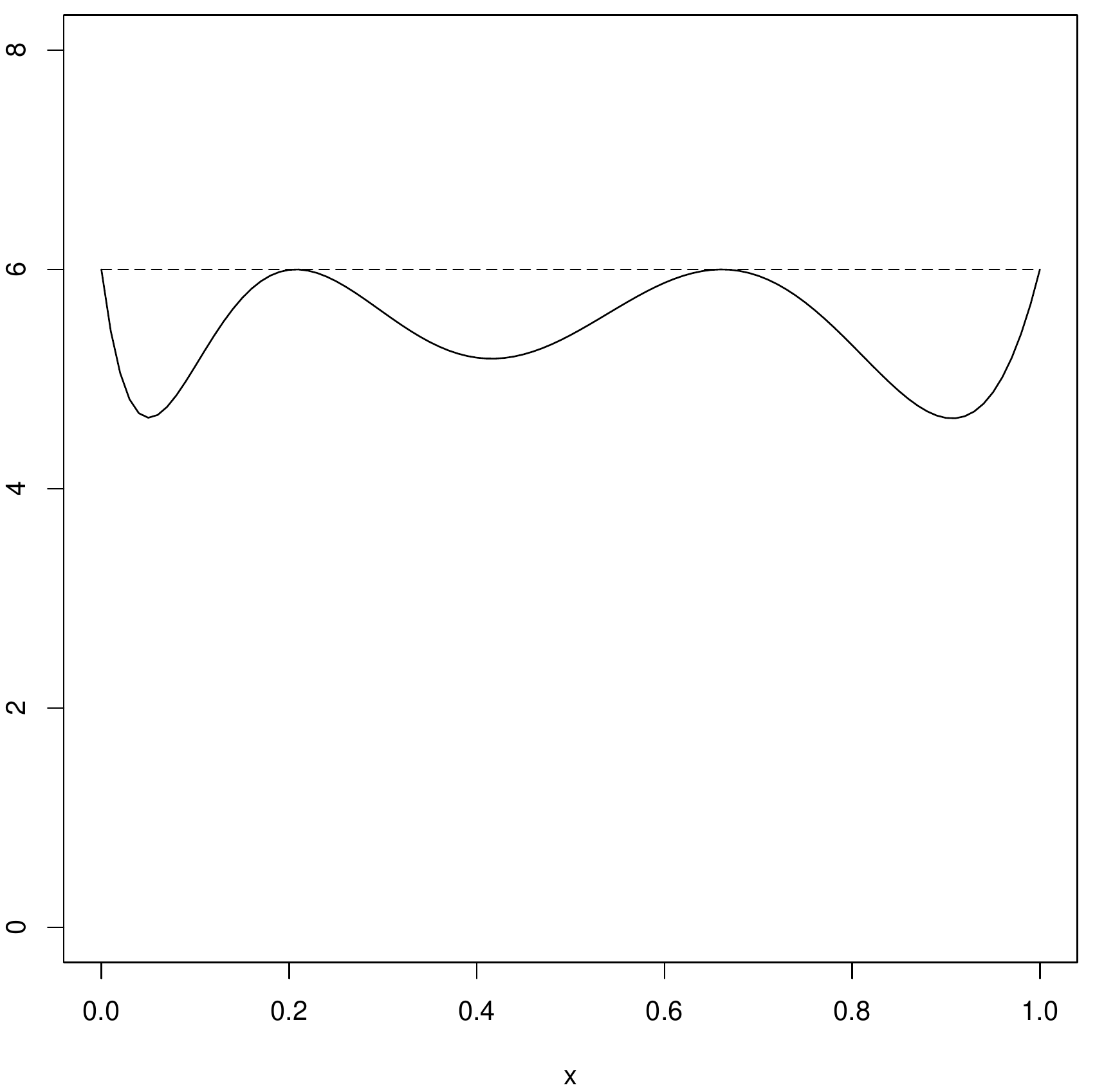}~~
      \includegraphics[width=5cm]{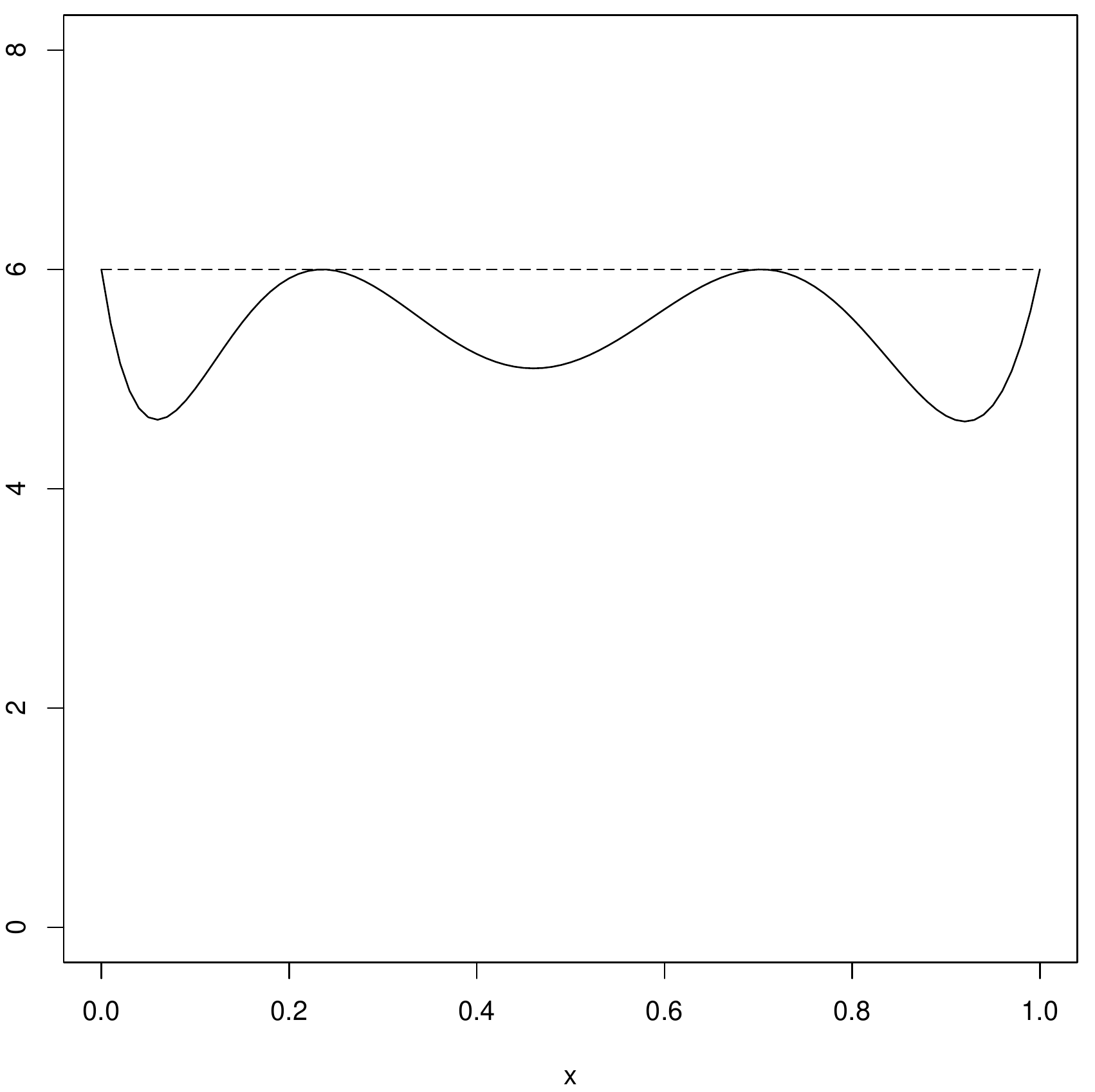} ~
            \includegraphics[width=5cm]{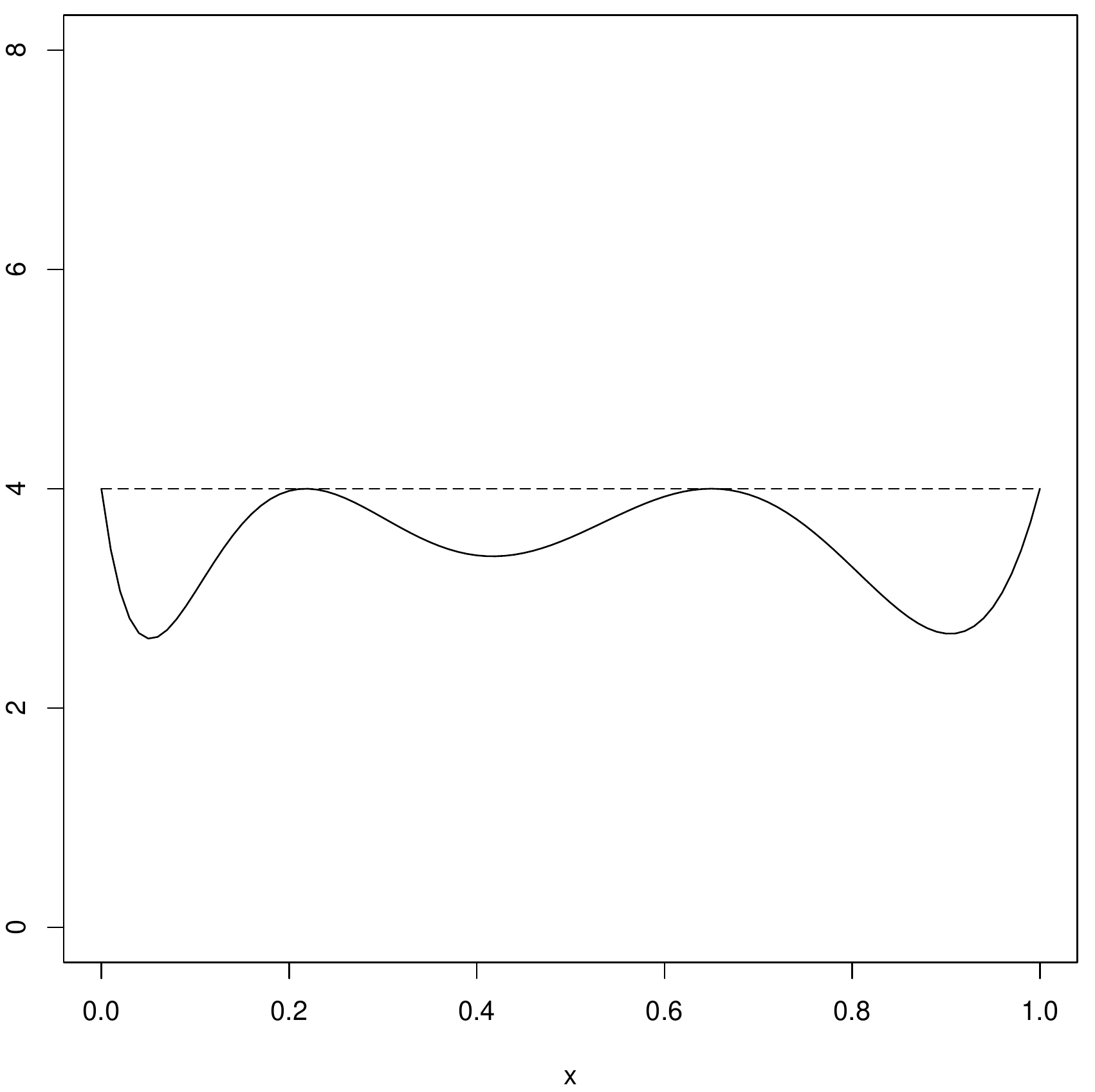}
    \caption{\label{fig1} \it The necessary condition of optimality for
three Bayesian-optimal $4$-point designs
    in the cubic polynomial regression  model with variance structure
\eqref{var2}. Left: Bayesian $D$-optimal design with respect to a uniform prior.
Middle: Bayesian
    optimal design with respect to the Jeffreys prior. Right:
    Bayesian  optimal design with respect to the Berger-Bernardo-prior. }
\end{figure}

 \begin{table}[h!]
\begin{center}
  \begin{tabular}{|c|c|c|c|c|c|}\hline
  \multicolumn{2}{|c|}{\eqref{bayd} with \eqref{puni}}&\multicolumn{2}{|c|}{\eqref{d8}}&\multicolumn{2}{|c|}{\eqref{BB}}\\\hline
  $\xi(x_i)$&$x_i$	&$\xi(x_i)$&$x_i$&$\xi(x_i)$&$x_i$		\\\hline\hline
  0.3356&0&0.3624&0&0.3333&0  \\\hline
  0.2686&0.6532&0.2527&1.1859&0.3333& 0.6340\\\hline
  0.3958&3&0.3849&3&0.3333&2.36603 \\\hline
   \end{tabular}
 \caption{ \label{tab2} \it Bayesian optimal $3$-point designs with respect to non-informative priors  for the
 quadratic polynomial regression model on the interval $[0,3]$ with variance structure  \eqref{var2}, where
 $\theta_{n+2} \in [0,4]$. Left column: Bayesian $D$-optimal designs with respect to the uniform prior. Middle column: Bayesian optimal designs
 with respect to the Jeffreys prior. Right  column: Bayesian optimal designs
 with respect to the Bernardo-Berger prior.}

 \end{center}
 \end{table}

  \begin{figure}[hbt!]
     \includegraphics[width=5cm]{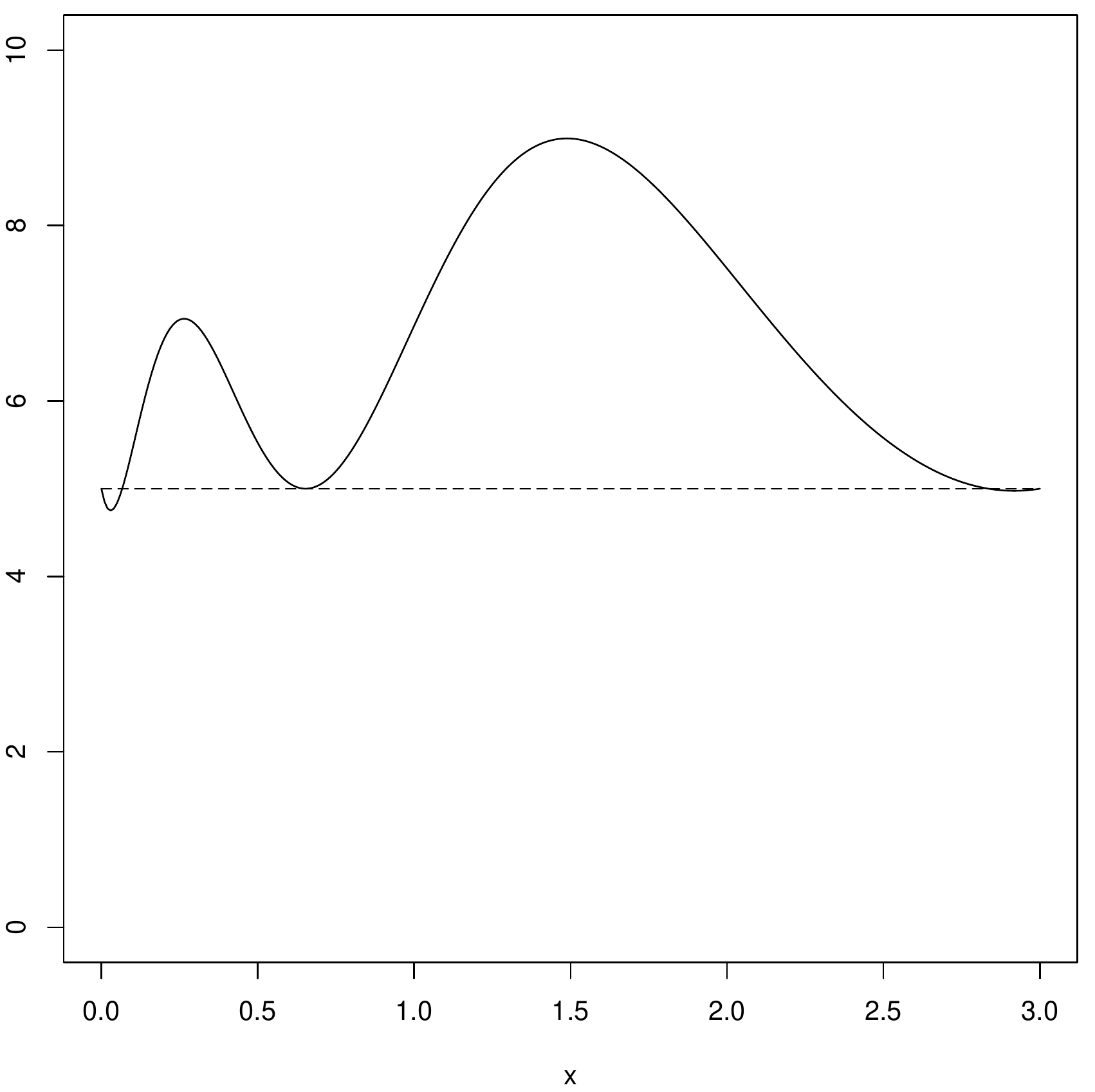}~~
      \includegraphics[width=5cm]{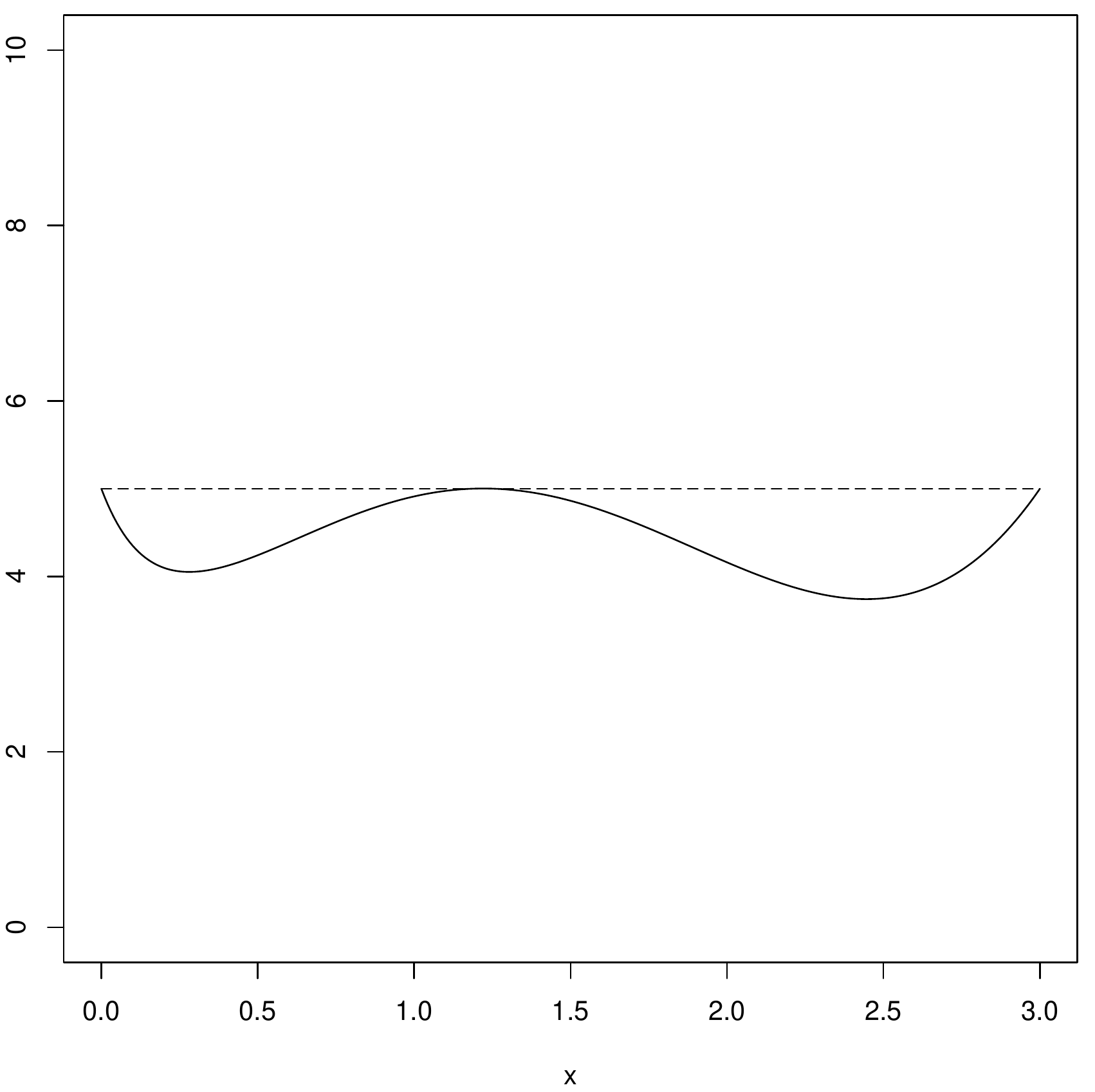} ~
           \includegraphics[width=5cm]{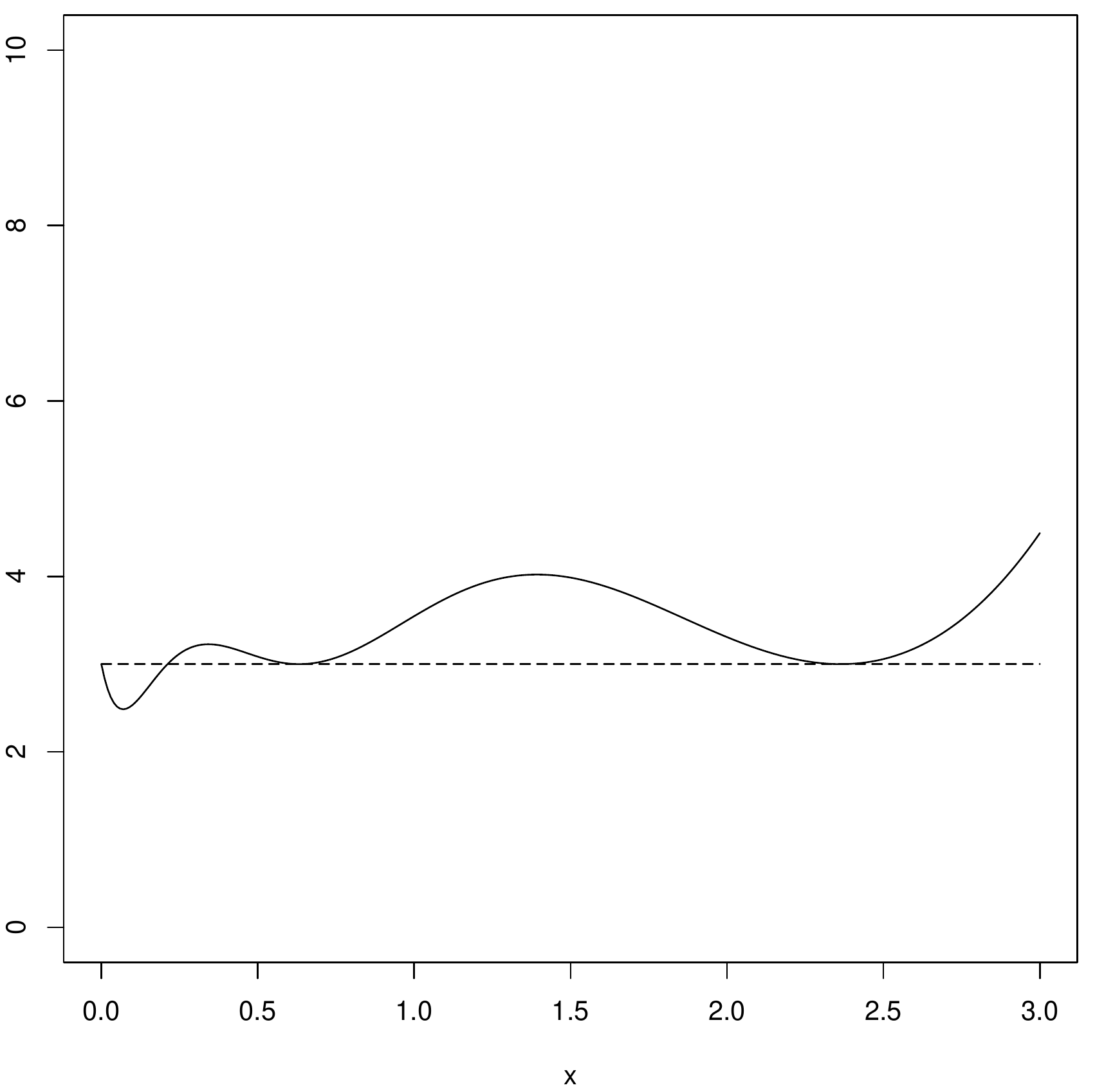}
    \caption{\label{fig1b} \it The necessary condition of optimality for
  Bayesian optimal $3$-point designs in the quadratic polynomial regression  model with variance structure
\eqref{var2}.
Left: Bayesian $D$-optimality.
Middle: Bayesian
    optimal design with respect to the Jeffreys prior.
 Right:
    Bayesian  optimal design with respect to the Berger-Bernardo-prior.}
\end{figure}

 \begin{table}[h!]
\begin{center}
  \begin{tabular}{|c|c|c|c|c|c|}\hline
  \multicolumn{2}{|c|}{\eqref{bayd} with \eqref{puni}}&\multicolumn{2}{|c|}{\eqref{d8}}&\multicolumn{2}{|c|}{\eqref{BB}}\\\hline
  $\xi(x_i)$&$x_i$	&$\xi(x_i)$&$x_i$&$\xi(x_i)$&$x_i$		\\\hline\hline
  0.3209&0&0.3624&0&0.3193&0  \\\hline
  0.1931&0.4480&0.2527&1.1859&0.2478& 0.4728\\\hline
  0.1601& 1.2939&0.3849&3& 0.2453&1.4472\\\hline
  0.3259&3& &&0.1876&3\\\hline
   \end{tabular}
 \caption{ \label{tab2b} \it Bayesian optimal $3$- or $4$- point designs with respect to non-informative priors  for the
 quadratic polynomial regression model on the interval $[0,3]$ with  variance structure  \eqref{var2}, where
 $\theta_{n+2} \in [0,4]$. Left column: Bayesian $D$-optimal designs with respect to the uniform prior. Middle column: Bayesian optimal designs
 with respect to the Jeffreys prior. Right  column: Bayesian optimal designs
 with respect to the Bernardo-Berger prior.}
  \end{center}
 \end{table}}
\end{example}

We finally briefly discuss optimal designs with respect the variance function  \eqref{var1}. In this case we are only able to
determine the Bayesian optimal designs with respect to the Berger-Bernardo  prior.

  \begin{theorem} \label{thm4} Consider the polynomial regression model \eqref{pol}.
If the design space and the variance function are given by ${\cal X}=(-1,1)$  and by \eqref{var1}, respectively, then
   the Bayesian optimal $(n+1)$-design with respect to the Berger-Bernardo prior  puts equal masses at the roots of the $(n+1)$th Jacobi polynomial
   $P_{n+1}^{(g_1,g_2)} (x)$, where the parameters $g_1$ and $g_2$ are given by
   $$
   g_j   ={\int \theta_{n+j} d \bm{\theta_2}  \over \int  d \bm{\theta_2} }~;~~j=1,2.
   $$
  \end{theorem}

  {\bf Proof:} Observing the representation  \eqref{fishpol2}
  we obtain for the lower block in the Fisher information matrix $ I(x, \bm{\theta})$ the representation
 $$ I_{22} (x, \bm{\theta}) =\left(\begin{array}{cc}
\log^2(1-x)&\log(x+1)\log(1-x)\\
\log(x+1)\log(1-x)&\log^2(x+1)
\end{array}\right).
$$
 Therefore we have
  $$
 \frac {|M_{22}(\xi,\bm{\theta})|^{1/2}}{\int |M_{22}(\xi,\bm{\theta_1},\bm{t_2})|^{1/2}d\bm{t_2}}   = {1\over \int d\bm{\theta_2}} = {1 \over \alpha_2},
$$
where the last equality defines the constant $\alpha_2$ in  an obvious manner. Consequently, for  an $(n+1)$-point design with masses  $\xi(x_0), \ldots , \xi(x_n)$  at the points $x_0, \ldots , x_n$ the optimality criterion reduces to
 \begin{eqnarray*}
 \Phi_{BB}(\xi)&=&
  \alpha_1 \exp\Bigl(\int\frac{1}{2\alpha_2} \log\Bigl[\prod_{j=0}^{n} \xi(x_j)\prod_{j=0}^{n}(1-x_j)^{\theta_{n+1}+1}(1+x_j)^{\theta_{n+2}+1}\prod_{\substack{
   m,\ell=0,...,n\\
   m<\ell
  }}(x_m-x_\ell)^2 \Bigr] d {\bm{\theta_{2}}} \Bigr)  \\
 &=&   \alpha_1 \exp\Bigl(\int\frac{1}{2\alpha_2} \log \Bigl[ \prod_{j=0}^{n }(1-x_j)^{\theta_{n+1}+1}(1+x_j)^{\theta_{n+2}+1}
 |H_n (\xi )| \Bigr] d {\bm{\theta_{2}}}\Bigr),
  \end{eqnarray*}
  where  $\alpha_1:=\int d{\bm{\theta_1}}$ and the matrix $ H_n (\xi )$ is the Hankel matrix of the $(n+1)$-point design $\xi$, that is
  $$
H_n (\xi) = (c_{i+j}(\xi) )_{i,j=0,\ldots n} = \prod^n_{j=0} \xi (x_j) \prod_{\substack{
   m,\ell=0,...,n\\
   m<\ell
  }} (x_m - x_{\ell})^2.
   $$
   Observing Theorem \ref{thm2} it therefore follows that the Bayesian $(n+1)$-point optimal design with respect to the Bernardo-Berger prior
   can be determined by maximizing the expression
   $$
   \Bigl(  \prod_{i=1}^{2n+1}q_{i}  \Bigr) ^{ g_1+1}
\Bigl( p_{2n+1}\prod_{i=1}^n (p_{2i-1}q_{2i}) \Bigr) ^{ g_2 +1 }
 \prod_{i=1}^n (q_{2i-2}p_{2i-1}q_{2i-1}p_{2i})^{n-i+1}
   $$
   with respect to the canonical moments $p_1, \ldots , p_{2n+1}$. Straightforward algebra gives for the corresponding ``optimal" canonical moments
\begin{eqnarray*}
   p_{2i-1} &=& { g_2  + n+1-i \over g_1 +g_2 + 2(n+1-i)}
   ~; ~~i=1,\ldots , n+1 \\
     p_{2i} &=& { n+1-i \over g_1  +g_2 + 2(n+1-i)+1}
     ~; ~~i=1,\ldots , n+1.
     \end{eqnarray*}
     The design corresponding to these canonical moments has been determined in \cite{studden1982a} and puts equal masses at the roots
     of the $(n+1)$th Jacobi polynomial $P_{n+1} ^{ (g_1 ,g_2) } (x) $ [see also \cite{dettstud1997} for an alternative proof], which completes the proof of Theorem \ref{thm4}.
\hfill $\Box$

\section{Bayesian optimal designs  for nonlinear regression} \label{sec5}
\def\theequation{5.\arabic{equation}}
\setcounter{equation}{0}

In this section we illustrate the application of the methodology determining Bayesian optimal designs for the
 EMAX model and a compartment model, which are frequently used  in pharmacology. {Locally optimal designs for this model have been determined by numerous authors [see
 \cite{atkinson1993},  \cite{jones1999}, \cite{debrpepi2008}  and \cite{detkisbevbre2010}] and we present some Bayesian optimal designs with respect to non-informative priors.

For both models we assume that the response at experimental condition $x \in \mathcal{X}$ is normally distributed with mean $\mu(x,\bm{\theta})$ and variance $\sigma^2(\bm{\theta})=\theta_3>0$. Here the variance is considered as a nuissance parameter. For the criterion \eqref{bayd} we use a
uniform and a  functional uniform prior for the parameters $(\theta_0, \theta_1, \theta_2)$ and {an arbitrary prior for $\theta_3$}. The criteria with respect to the Jeffreys prior and the Berger-Bernardo-prior are equivalent in this case. All designs have been calculated numerically using Maple.  }

We begin with the EMAX model which describes a dose-response relationship
$$
\mu(x,\bm{\theta})=\theta_0+\frac{\theta_1x}{x+\theta_2},
$$
where $\theta_1$ 	determines the asymptotic maximum effect, $\theta_2$ the dose that gives half of the asymptotic maximum effect and $\theta_0$ describes the effect of placebo.

In Table \ref{tab3} we display some Bayesian optimal designs with respect to non-informative priors, where the design space is given by the interval $[0,4]$. For the parameters we assume $\theta_0\ge 0, \theta_1 \in (0,5]$, and $\theta_2 \in [1,6]$, where $\theta_0$ and $\theta_3$ are each from a compact interval.

 \begin{table}[h!]
\begin{center}
  \begin{tabular}{|c|c|c|c|c|c|}\hline
  \multicolumn{2}{|c|}{\eqref{bayd} with \eqref{puni}}&\multicolumn{2}{|c|}{\eqref{d8}/\eqref{BB}}&\multicolumn{2}{|c|}{\eqref{bayd} with \eqref{fununi}}\\\hline
    $\xi(x_i)$&$x_i$    &$\xi(x_i)$&$x_i$&$\xi(x_i)$&$x_i$            \\\hline\hline
  0.333&0&0.333&0&0.333& 0 \\\hline
  0.333 &1.2028&0.333&0.9472&0.333&0.9766\\\hline
  0.333&4&0.333&4 &0.333&4\\\hline
  \end{tabular}
 \caption{ \label{tab3} \it Bayesian optimal $3$-point designs with respect to non-informative priors  for the
 EMAX model on the interval $[0,4]$. Left column: Bayesian $D$-optimal design with respect to the uniform prior. Middle  column: Bayesian optimal designs
 with respect to the Jeffreys and   Bayesian optimal designs
 with respect to the Bernardo-Berger prior. Right column: Bayesian $D$-optimal designs with respect to the functional uniform prior.}

 \end{center}
 \end{table}

\begin{figure}[hbt!]
\begin{center}
      \includegraphics[width=5cm]{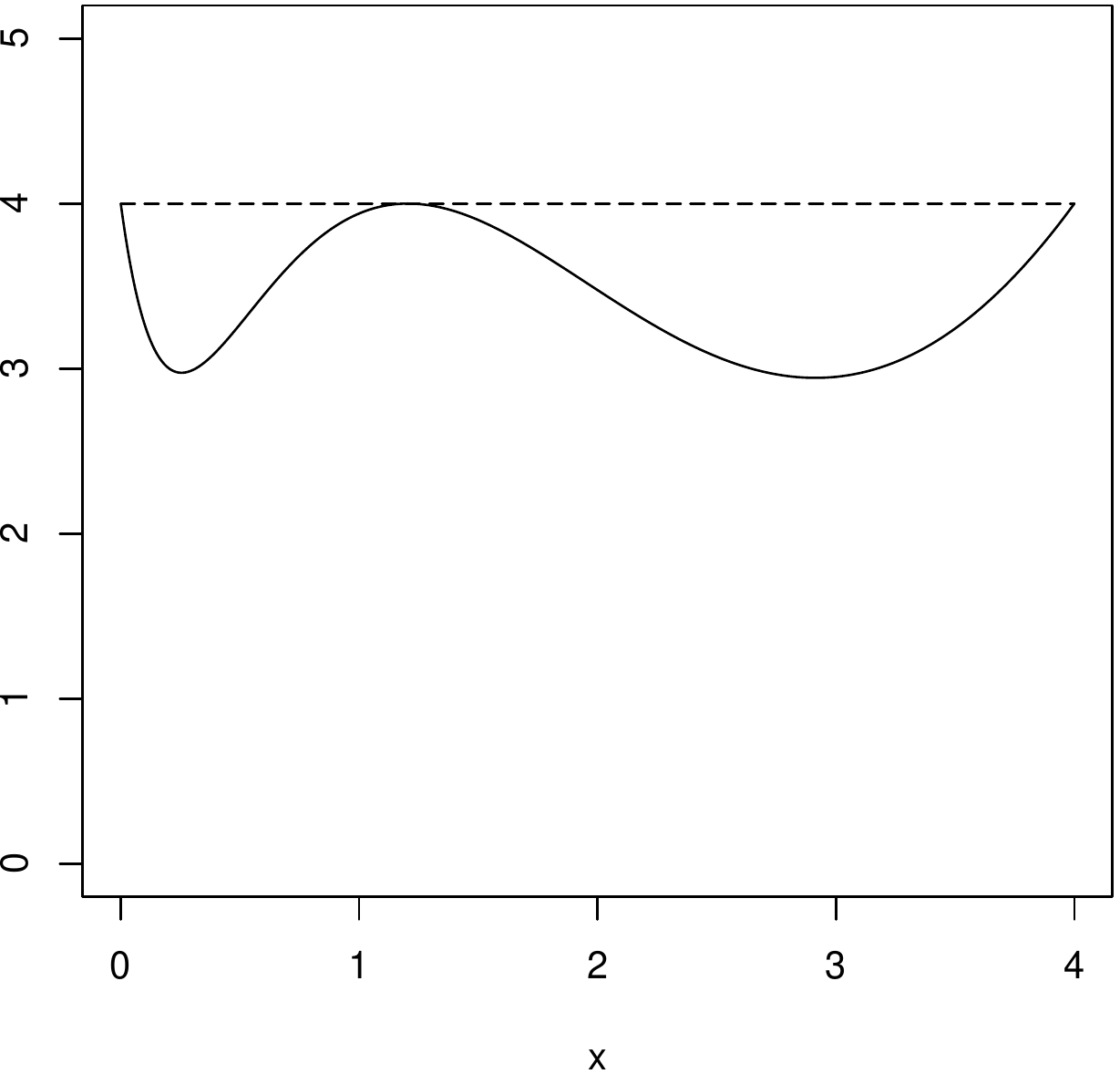}
  ~
           \includegraphics[width=5cm]{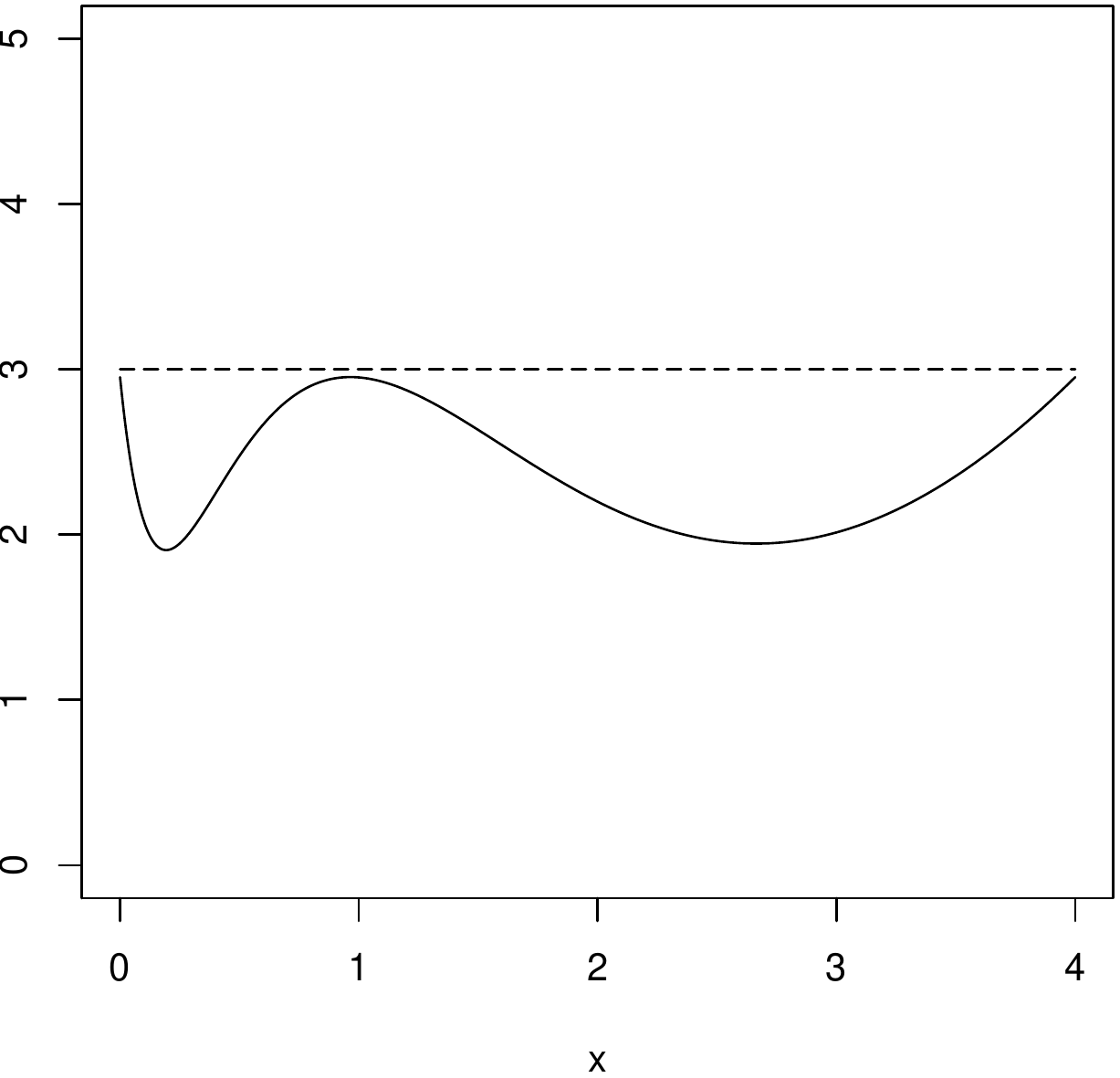}
  ~
      \includegraphics[width=5cm]{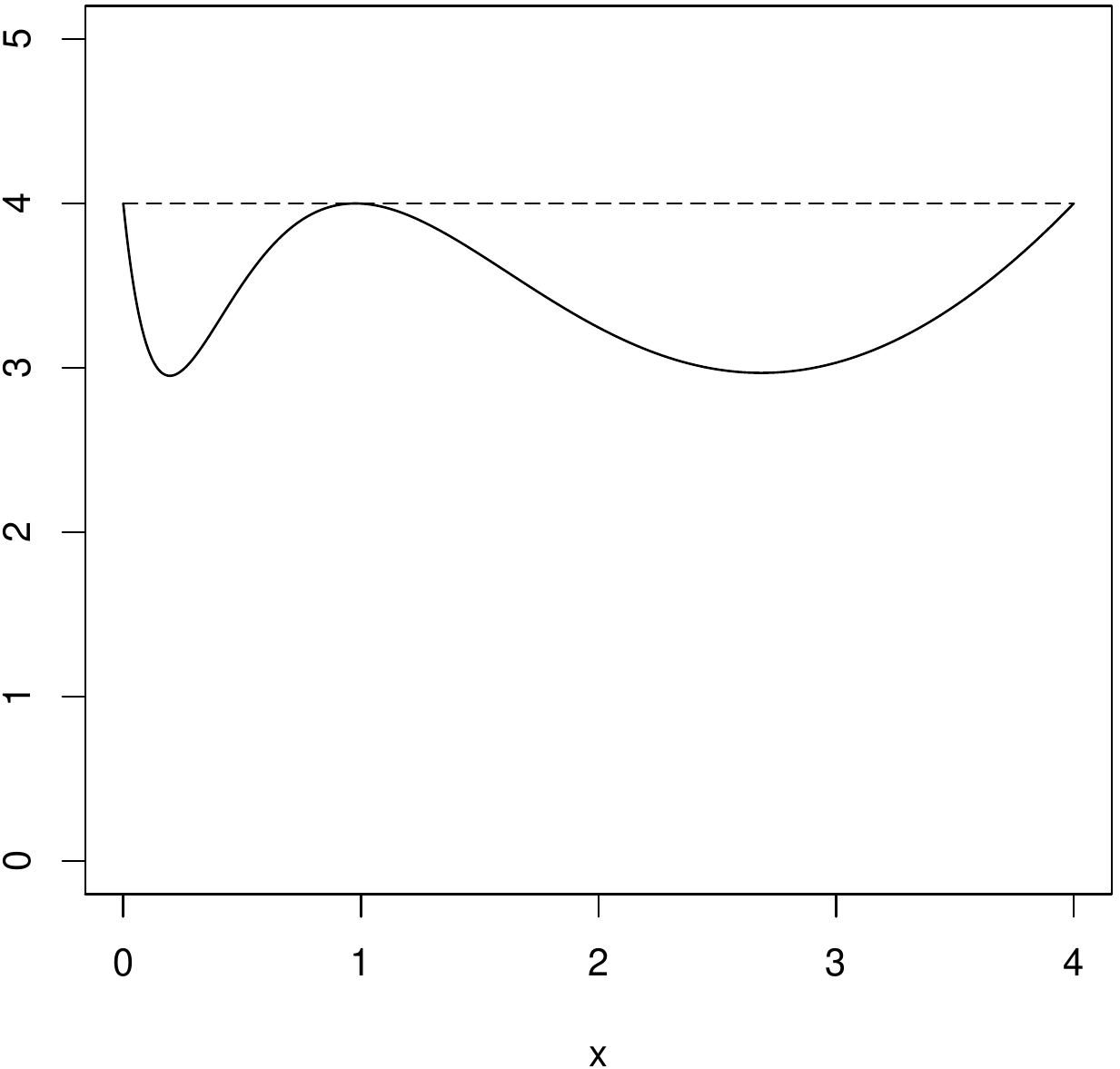}
    \end{center}
    \caption{\label{fig2} \it The necessary condition of optimality for optimal $3$-point designs
    in the EMAX model. Left: Bayesian $D$-optimality
design with respect to a uniform prior.  Middle: Bayesian
    optimal design with respect to the Jeffreys prior  and the Berger-Bernardo-prior. Right: Bayesian optimal design
with respect to a functional uniform prior.}
\end{figure}

We observe that the Bayesian-optimal $3$-point designs with respect to the Jeffreys prior and the Berger-Bernardo prior and the Bayesian D-optimal design with respect to the functional uniform prior look similar, while the Bayesian $D$-optimal design with respect to the uniform prior has a larger interior  support point. The application of Theorem \ref{thmfunct}
- \ref{thmBB} is illustrated in Figure \ref{fig2}. We observe that all designs satisfy the necessary condition for optimality.

 We conclude this paper with a brief discussion of Bayesian optimal designs for a compartment model, which is used as a model for   the concentration of a substrate over time   involving   absorption and the elimination of a substrate. Here the mean is given by
$$
\mu(x,\bm{\theta})=\theta_0 (\exp(-\theta_1x)- \exp(-\theta_2x)),
$$
where
 $\theta_1$ is the elimination constant and  $\theta_2$ the absorption constant.  The corresponding optimal designs are displayed in Table \ref{tab5}, where the design space is given by $\mathcal{X} = [0,20]$ and
 $\theta_0 >0, \theta_1 \in [0.05, 0.07]$, and $ \theta_2 \in [3.3, 5.3]$ [see  \cite{atkinson1993}]. As before $\theta_0$ and $\theta_3$ are each from a compact interval. All designs presented in this table satisfy the necessary condition of optimality (the corresponding plots are not displayed for the sake of brevity). Interestingly all designs exhibit a very similar structure.

 \begin{table}[h!]
\begin{center}
  \begin{tabular}{|c|c|c|c|c|c|c}\hline
  \multicolumn{2}{|c|}{\eqref{bayd} with \eqref{puni}}&\multicolumn{2}{|c|}{\eqref{d8}/\eqref{BB}}&\multicolumn{2}{|c|}{\eqref{bayd} with \eqref{fununi}}\\\hline
    $\xi(x_i)$&$x_i$    &$\xi(x_i)$&$x_i$&$\xi(x_i)$&$x_i$            \\\hline\hline
  0.333&0.2286&0.333&0.2321&0.333&0.2343 \\\hline
  0.333&1.4106&0.333&1.4310&0.333&1.4420\\\hline
  0.333&18.1145&0.333&18.3185&0.333&18.3132\\\hline
  \end{tabular}
 \caption{ \label{tab5} \it Bayesian optimal $3$-point designs with respect to non-informative priors  for the
 compartment model. Left column: Bayesian $D$-optimal designs with respect to the uniform prior. Middle  column: Bayesian optimal designs
 with respect to the Jeffreys and the Bernardo-Berger prior. Right column: Bayesian $D$-optimal designs with respect  to the functional uniform prior.}
 \end{center}
 \end{table}
\bigskip
\bigskip

{\bf Acknowledgements.} The authors would like to thank Martina Stein, who typed parts of this manuscript with considerable technical expertise.
This work has been supported in part by the Collaborative
Research Center ``Statistical modeling of nonlinear dynamic processes'' (SFB 823, Teilprojekt C2) of the German Research Foundation (DFG).

\bibliographystyle{apalike}

\bibliography{burg_dett}

\end{document}